# Electronic properties of zigzag, armchair and their hybrid quantum dots of graphene and boron-nitride with and without substitution: A DFT study

Sharma SRKC Yamijala [1, #], Arkamita Bandhyopadyay [2, #] and Swapan K Pati [2, 3, *]

[1] Chemistry and Physics of Materials Unit, Jawaharlal Nehru Centre for Advanced Scientific Research, Jakkur, Bangalore, Karnataka 560064, India
[2] New Chemistry Unit, Jawaharlal Nehru Centre for Advanced Scientific Research, Jakkur, Bangalore, Karnataka 560064, India
[3] Theoretical Sciences Unit, Jawaharlal Nehru Centre for Advanced Scientific Research, Jakkur, Bangalore, Karnataka 560064, India
# These authors have contributed equally to this work
* Corresponding Author, E-mail: pati@jncasr.ac.in.

## Abstract

Spin-polarized density functional theory calculations have been performed on armchair graphene quantum dots and boron-nitride quantum dots (A-G/BN-QDs) and the effect of carbon/boron-nitride substitution on the electronic properties of these A-G/BN-QDs has been investigated. As a first step to consider more realistic quantum dots, quantum dots which are a combination of zigzag QDs and armchair QDs have been considered. Effect of substitution on these hybrid quantum dots has been explored for both GQDs and BNQDs and such results have been compared and contrasted with the results of substituted A-G/BN-QDs and their zigzag analogues. Our work suggests that the edge substitution can play an important tool while tuning the electronic properties of quantum dots.

## Introduction

Complex devices based on graphene nanoribbons (GNRs), like, GNR-FETs,[1] p-n junctions,[2] spin-filters[3] etc. have been studied theoretically and some of them have already been realized experimentally.[4] Recently, researchers have started focusing on various shaped nanoribbon junctions which could be the plausible building blocks for the 2D-nano-networks. Several shaped nanoribbon junctions namely, L-shaped,[5] T-shaped,[6] cross-shaped[7-9], S-shaped[10] and Z-shaped[11] nanoribbon junctions have been studied and most of these studies have concentrated on the conduction properties of these junctions.

A Z-shaped GNR junction has been shown as a promising candidate to confine the electronic states completely i.e. a quantum dot (QD) can be realized at the junction.[11] Also, conductance through a Z-shaped nano-ribbon junction is highly dependent on the angle and the width of the junction.12 Not only GNR, but also boron-nitride nanoribbon (BNNR) junctions have been studied and Z-shaped BNNR junctions were shown to exhibit spin-filtering as well as rectifying effects at the nano-junction depending upon the nature of the edge passivation.13 L-shaped GNRs show low reflectance to the electrons for a large included angle and high reflectance for low included angle when the L-shaped-junction is made of an armchair GNR (AGNR) and a zigzag GNR (ZGNR), and, an opposite effect has been observed when the L-shaped-junction is made of two ZGNRs.5 Similar spin-polarized calculations on the in-plane conductance of the GNR-cross points at different angles have shown large-scattering for quantum transport, except when two ZGNR ribbons meet at 60° angle.14 Studies on the T-shaped junctions showed that, these systems are metallic and their conduction properties are sensitive to the height of the stem and the doping position, i.e. on the stem or on the shoulder.6 Similar results were reported for the cross-shaped ribbons. Spin-polarized-conductance calculations on a cross-shaped junction show a transverse spin current with zero charge-current.7

Motivated by the above mentioned interesting spin-polarized conducting properties of cross-shaped ribbon networks obtained from the theoretical calculations7-9 and by the recent experimental realization of the alphabetical character QDs,15 we have performed the spin-polarized density functional theory (DFT) calculations to understand the electronic and magnetic properties of the cross-shaped ('+' shaped) graphene (G) and boron-nitride (BN) QDs. These QDs can be considered as the low-dimensional siblings of GNRs and can be visualized as the QDs formed by the intersection of a zigzag GQD (ZGQD) and an armchair GQD (AGQD) as shown in figure 2. Earlier works on GQDs suggest that substitution on GQDs (with boron/nitrogen) and on BNQDs (with carbon) show quite interesting properties, such as, spin polarized energy gap, wide range of optical absorption etc.16 Inspired by these interesting findings, we have also studied the substitutional effects on the electronic and magnetic properties of the cross-shaped GQDs and BNQDs. Also, experimentally hybrid BNC sheets have already been prepared, where the ratio of C:BN17 and the shape or size of BN domain on C or vice-versa can be controlled precisely.18 Thus, easily these substituted QDs can be experimentally synthesized using the conventional lithography19-22 or the more recent "nanotomy" technique.21

As cross-shaped shaped graphene (G) and boron-nitride (BN) quantum dots (QDs) can be seen as a hybrid (or combination) of zigzag QDs (ZQDs) and armchair QDs (AQDs) of equal length and width, first we have performed a series of calculations on both substituted and pristine armchair QDs (AQDs). Such calculations are important to understand the electronic properties of cross-shaped QDs and, to the best of our knowledge, such calculations have not been performed earlier. Next, we have investigated the electronic and magnetic properties of the hybrids of zigzag and armchair QDs (HQDs) of various sizes and different levels of substitution. It is known that, the electronic and magnetic properties of these types of QDs are highly dependent on the nature of the edges,23-25 and hence, we have considered the substitution of these pristine QDs only at the edges. For a GQD, edges are substituted with boron and nitrogen (BN) pairs and for a BNQD, edges are substituted with carbon. Thus, the systems are iso-electronic before and after the substitution. We have passivated all the systems with H-atoms, because in several previous studies, have shown that H passivation makes the systems more stable compared to the corresponding pristine QDs. Depending on whether all the edge atoms are substituted or only half of them are substituted, we named these QDs as completed-ed-QDs or partial-ed-QDs, respectively, as shown in the figures 1 and 2.

## Computational Details:

All the first-principles calculations have been performed using spin-polarized density functional theory (DFT) as implemented in the SIESTA package.28 The generalized gradient approximation (GGA) with the Perdew-Burke-Ernzerhof (PBE) form29 is chosen for the exchange-correlation functional. Interaction between the ionic cores and the valence electrons is accounted by the norm conserving pseudo-potentials in the fully nonlocal Kleinman-Bylander form.30 The pseudo-potentials are constructed from 3, 4 and 5 valence electrons for the B, C and N atoms, respectively. To expand the wave-functions, numerical localized combination of atomic orbitals with double-$\zeta$ basis sets are used. To represent the charge density, a reasonable mesh-cut-off of 400 Ry is used for the grid integration. As the systems under consideration are quantum dots, only gamma point of the Brillouin-zone has been considered for both the electronic property calculations and for the full-relaxation of the systems. Systems are considered to be optimized only when the forces acting on all the atoms are less than 0.01 eV / Å. A vacuum of 20 Å has been maintained in all the three direction surrounding the QDs to avoid any unwanted interactions between the quantum-dots and their periodic images. For all the DOS and pDOS plots, a broadening parameter of 0.025 eV has been used.

## Results and Discussion:

### *ZQDs and AQDs:*

We consider all the systems as spin-polarized and carried out extensive DFT calculations. However, we find that in these systems there is no spin-polarization. Thus, we are reporting all our results without any spin-polarization unless otherwise stated. Earlier works16 predicted zigzag edged rectangular QDs are semiconducting and up on substitution with BN pairs, ZGQDs have shown interesting electronic and optical properties. We have performed a series of calculations on both substituted and pristine armchair QDs (AQDs). Results of AQDs have been compared and contrasted with the results of ZQDs reported earlier.[16] For this, we have considered the systems of medium sizes i.e. (33, 4). We have chosen only one size as changes in the electronic properties with the system size are found to be negligible.[16]

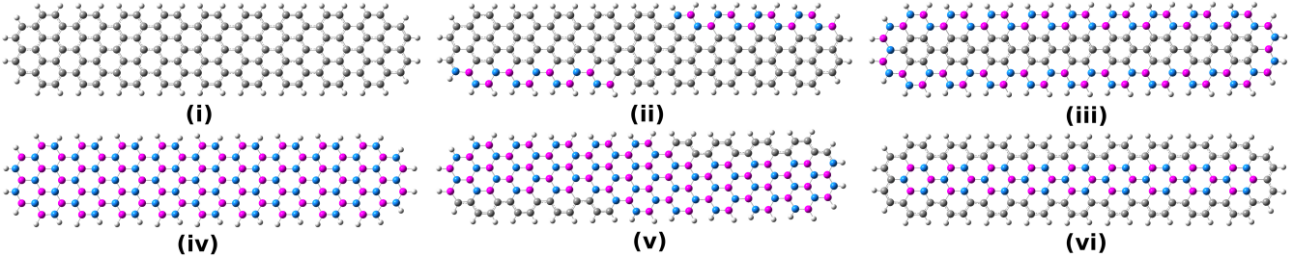

Figure 1: Optimized geometries of (i) AGQD, (ii) BN-partial-ed-AGQD, (iii) BN-full-ed-AGQD, (iv) ABNQD, (v) C-partial-ed-ABNQD and (vi) C-full-ed-ABNQD

As shown in table 1, we find that (a) DOS near the Fermi-level of AQDs is also mainly due to the carbon atoms (whenever carbon atoms are present in the system) and (b) their HOMO-LUMO gap (HLG) can be tuned through substitution as were found in case of ZQDs.[16] One notable result is the huge decrement in the HLG for both the complete and partial edge substituted BNQDs when compared to the pristine BNQD. But, all the (33, 4) AQDs are semi-conducting and, unlike their ZQD counter parts, none of the AQDs are spin-polarized near the Fermi-level (even after substituting their edges with BN-pairs either partially or completely). Thus, AQDs' electronic properties can be considered as more rigid towards substitution compared to the ZQDs. As real QDs (experimentally obtained ones), can contain both zigzag and armchair edges/wings, we have also considered a mixture of armchair and zigzag wings in our system.

Table 1: Formation energy and HLG values of all the AQDs

| System Name | Formation Energy(eV) | H-L Gap (eV) |
| --- | --- | --- |
| AGQD | -0.555 | 1.62 |
| BN-partial-ed- AGQD | -0.475 | 0.55 |
| BN-full-ed- AGQD | -0.401 | 2.06 |
| ABNQD | -0.498 | 4.41 |
| C-partial-ed-ABNQD | -0.443 | 0.91 |
| C-full-ed-ABNQD | -0.462 | 0.81 |

*HQDs:*

As there will be several possibilities while mixing the two types of wings, we have considered a simple case, namely, the joining of a ZQD with an AQD to form a '+' shaped QD. In order to see the size effects of the wings on the calculated properties, we have considered 3 different sizes of the wings. At the same time, to avoid any confusion in understanding the new findings due to substitution (a) we kept the central/junction region of all the QDs as same (for all the 3 sizes) [which helped us to rule out (automatically) the width dependent electronic properties of AGQD (i.e. 3p, 3p+1 and 3p+2, where 'p' is an integer[31])] and (b) in all the 3 sizes, we have maintained equal numbers of zigzag and armchair edges on both sides of the junction [which helped us to keep track on the important changes which will occur only because of the substitutional effects rather than the asymmetry across the junction]. The size of the QD is indicated by (A, Z), where 'A' and 'Z' are the lengths of the armchair and zigzag-edged-wings, respectively, in nm and the 3 sizes considered are (2.65, 2.67), (4.39, 4.16) and (6.12, 5.64). All the six systems of the medium size, (4.39, 4.16), are shown in figure 2 and all of them have a total number of 240 atoms (308 atoms by including the hydrogen atoms), among which, 130 atoms are always at the edges of the QD and the remaining 110 atoms at the center, which makes these systems suitable for studying the edge effects. We believe that, our results will give the preliminary understanding of these HQDs (hybrids of ZQDs and AQDs).

## Stability:

As HQDs have armchair and zigzag wings, all the 3 different spin-configurations, namely, ferromagnetic (FM), anti-ferromagnetic (AFM) and nonmagnetic (NM) configurations have been considered in this study. Both the formation energy and relative energy (with respect to the energy of the most stable spin-configuration) of all these configurations are given in the table S1 of SI. As shown in the table S1, all the systems are thermodynamically stable (i.e. their formation energy values are negative), and hence, in principle, they should be experimentally realizable (based on the techniques discussed in introduction). Indeed, similar QDs (or more precisely, ribbons as lengths are of ~10 nm) have already been realized experimentally in the shapes of X (also can be seen as the HQDs of present study), Y and Z.[15] Among all the QDs, we find that, only HGQD and BN-partial-ed-HGQD have an AFM ground state. For all other HQDs, the difference between the energies among different spin-configurations is either below the room-temperature or very near the room-temperature (see table S1), and hence, we considered the ground states of these systems as non-magnetic only. Thus, our calculations suggest that all the HQDs can be synthesized experimentally and they will be non-spin-polarized in their ground states. Next, we will discuss the electronic properties of these HQDs (in their ground states) and compare them with the electronic properties of the respective ZQDs and AQDs from which these HQDs are formed.

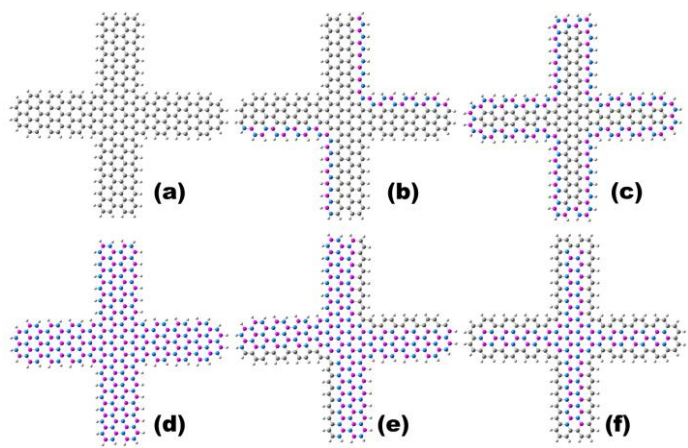

Figure 2: Optimized geometries of (a) HGQD, (b) BN-partial-ed-HGQD, (c) BN-full-ed-HGQD, (d) HBNQD, (e) C-partial-ed-HBNQD and (f) C-full-ed-HBNQD

**Electronic properties:**

To understand the electronic properties of HQDs, we have calculated the HOMO-LUMO gaps (HLGs) of each of these systems as shown in table S1. Here, we will discuss only the trends in the HLGs of the most stable spin-configurations of different HQDs. From the HLGs (see table S1) we find that the electronic properties of HQDs are indeed a mixture of the electronic properties of ZQDs and AQDs. For example, from table S1, we find that

  a) HLG of either HGQD or HBNQD has decreased with an increase in the system size as observed previously for ZQDs.16 Reason for the decrement is due to both more bulk like behavior (with an increase in size) and extended conjugation (for carbon containing systems).16
  b) For a particular size of HQD, partial substitution at the edges has shown a drastic decrement in the HLG values for both HGQD and HBNQD. For the case of HGQD, the decrement in the HLG after partial substitution is always > 0.2 eV and for HBNQD it is always > 4.0 eV (similar to both AQDs and ZQDs)
  c) Complete edge substitution of HGQDs with BN pairs increased their HLG by a maximum amount of 0.15 eV (this is similar to AQDs, see table 1, but, opposite of ZQDs), but for

HBNQDs, complete edge substitution with carbon atoms decreases its HLG by, at least, 3.4 eV (similar to both AQDs and ZQDs).16 The reason for the latter is because of the extended conjugation obtained by 'C' atom substitutions.

d) BN-partial-ed-HGQDs show a spin-polarized HLG in their ground state spin-configuration (similar to ZQDs, but, opposite of AQDs).

Thus, substitution can help in tuning the HLG of HQDs and the change in HLG is huge when the edges are substituted partially and the changes are highly prominent when the systems are HBNQDs (than HGQDs). To further understand these changes in the HLGs, we have plotted both the density of states (DOS) and projected DOS (pDOS) for all the HQDs.

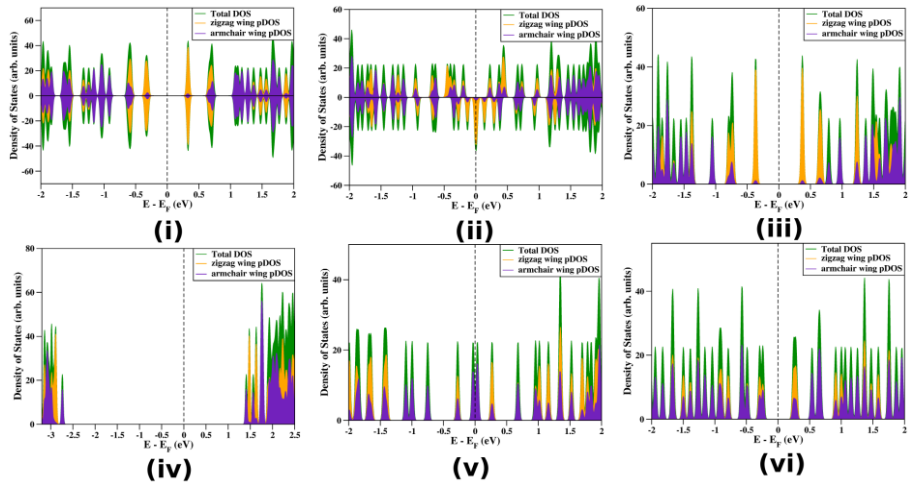

Figure 3: DOS and pDOS plots of (i) HGQD, (ii) BN-partial-ed-HGQD, (iii) BN-full-ed-HGQD, (iv) HBNQD, (v) C-partial-ed-HBNQD and (vi) C-full-ed-HBNQD. Green, orange and indigo colors indicate total DOS, zigzag wing pDOS and armchair wing pDOS, respectively.

In figure 3, we have given the DOS and pDOS of armchair and zigzag wings for all the HQDs of medium size [i.e. (4.39, 4.16)] in their respective ground states (see figure S2 for all the other DOS plots). Clearly, all the HQDs, except HBNQD and C-parital-ed-HBNQD, have greater zigzag edge contribution near the Fermi-level than the armchair edge. As the low-energy electronic properties are mainly dictated by the levels above and below (i.e. HOMO and LUMO) the Fermi-level, one should expect HBNQD and C-parital-ed-HBNQD (others) should have more AQD (ZQD) character than ZQD (AQD). Indeed, this is what we find in the DOS of the HQDs (compare figures 3 and S1). An interesting example where one can appreciate that "the electronic nature of a HQD is directly governed by the wing nature of the A/ZQD" is the BN-partial-ed-HGQD. As can be seen in figure 3(ii), BN-partial-ed-HGQD is a spin-polarized semi-conductor and the DOS and pDOS near the Fermi-level also compares exactly with that of BN-partial-ed-ZGQDs (see figure

4(i)). Figure 4 (ii-iv) shows the pDOS plots of the BN-partial-ed-HGQD of all sizes and all of them are spin-polarized semi-conductors, whose HOMOs and LUMOs are mainly composed of the states from zigzag wing C atoms. Similarly, C-partial-ed-BNQD's electronic structure is mainly dictated by the armchair wing C atoms as shown in figure 3e.

Thus, from this study we were able to understand that the electronic properties of HQDs are mainly dictated either by the electronic properties of the ZQD wing or the AQD wing of which a HQD is composed of.

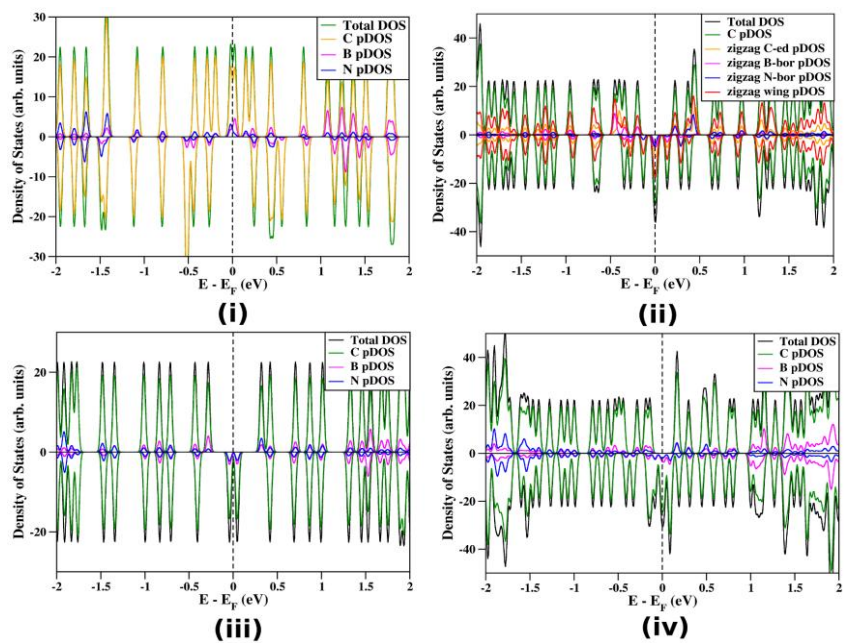

Figure 4: pDOS plots of BN-partial-ed (i) (33,4) -ZGQD, (ii) (4.39, 4.16) -HGQD, (iii) (2.65, 2.67) –HGQD and (iv) (6.12, 5.64) –HGQD

## Conclusions:

In conclusion, we have performed a series of calculations on AQDs and HQDs with and without substitution. Such calculations revealed that substitution can act as a powerful tool to determine the electronic properties of QDs. It has been shown that partial edge substitution can give rise to interesting properties like spin-polarized HOMO-LUMO gaps in HGQDs. Also, we find that, when a system has carbon atoms, then all the states near the Fermi-level are always from carbon atoms, and hence, the major electronic properties will be governed by the nature of these carbon states. Among HGQDs and HBNQDs, substitution has huge effects on HOMO-LUMO gap of HBNQDs compared to HGQDs. Finally, we found that the electronic properties of a HQD are mainly dictated by the electronic properties of either the ZQD or AQD from which the HQD is composed of.

# Acknowledgements:

SRKCSY, AB and SKP acknowledge TUE-CMS, JNCASR for the computational facilities. SRKCSY would like to thank Dr. Parida, Regensburg University for helpful discussions. AB acknowledges UGC for JRF.

# Supporting Information for "Electronic properties of zigzag, armchair and their hybrid quantum dots of graphene and boron-nitride with and without substitution: A DFT study"

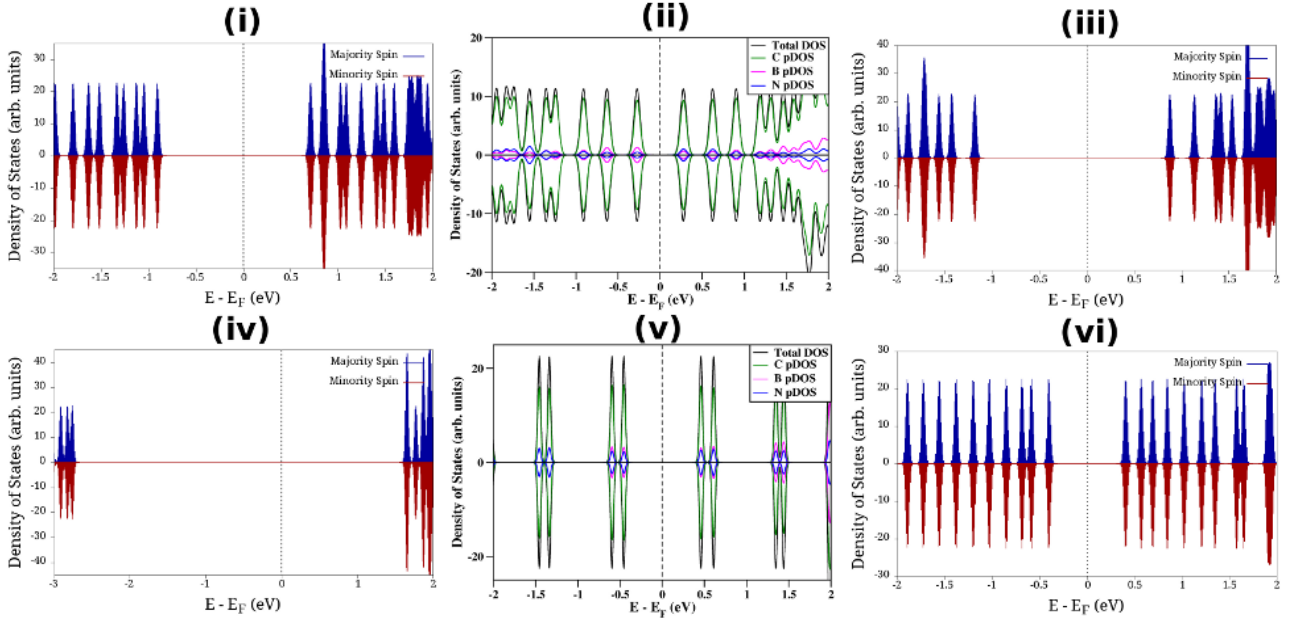

Figure S1: DOS and pDOS plots of (i) AGQD, (ii) BN-partial-ed-AGQD, (iii) BN-full-ed-AGQD, (iv) ABNQD, (v) C-partial-ed-ABNQD and (vi) C-full-ed-ABNQD

We have calculated the formation energy per atom, $E_{form}$, of all the systems in their ground state spin configurations using equation 1. The formation energy per atom, $E_{form}$, is defined as:

$$E_{Form} = [\,E_{tot} - (n_{CC} * \mu_{CC}) - (n_{BN} * \mu_{BN}) - (n_H * \mu_H)\,] / n_{tot} \quad \text{----}\,(1).$$

Here, $E_{tot}$ is the total energy of the system, $\mu_H$, $\mu_{BN}$ and $\mu_{CC}$ are the chemical potentials of a H atom (calculated from a hydrogen molecule), BN pair (calculated from an $8 \times 8$ h-BN supercell), and a CC pair (calculated from an $8 \times 8$ graphene super-cell), respectively and $n_H$, $n_{BN}$, $n_{CC}$ and $n_{Tot}$ are the number of H atoms, BN pairs, CC pairs and total number of atoms present in the system, respectively. In this calculation, the formation energies of graphene and BN-sheets are considered to be zero (i.e. they are considered to be reference states). More details of these calculations can be found in the references 1 and 2. [1,2]

Table S1: Spin polarization, formation energy, relative energy and H-L gap values for all the systems in all the 3 spin configurations. First column describes the system names, where the first index represents the type of system as described in the main paper (see figure 2). Second and third indices represent the no. of atoms in that system and the spin configuration, respectively.

| System Name | Spin- | Formation | Relative Energy | H-L gap |
|---|---|---|---|---|

| | polarization | Energy (eV/atom) | (eV) | spin-α | spin-β |
|---|---|---|---|---|---|
| HGQD_176_AFM | 0 | -0.509 | 0 | 0.92 | 0.92 |
| HGQD_176_FM | 0 | -0.509 | 0 | 0.92 | 0.92 |
| HGQD_176_NM | 0 | -0.509 | 0.03 | 0.92 | - |
| | | | | | |
| HGQD_308_AFM | 0 | -0.490 | 0 | 0.64 | 0.64 |
| HGQD_308_FM | 4.0 | -0.490 | 0.09 | 0.76 | 0.70 |
| HGQD_308_NM | 0 | -0.490 | 0.4 | 0.12 | - |
| | | | | | |
| HGQD_440_AFM | 0 | -0.482 | 0 | 0.64 | 0.64 |
| HGQD_440_FM | 4.0 | -0.481 | 0.27 | 0.43 | 0.34 |
| HGQD_440_NM | 0 | -0.480 | 0.713 | 0.0041 | - |
| | | | | | |
| BN_partial_ed_HGQD_176_AFM | 0 | -0.424 | 0 | 0.60 | 0.09 |
| BN_partial_ed_HGQD_176_FM | 0.8 | -0.424 | 0.01 | 0.27 | 0.30 |
| BN_partial_ed_HGQD_176_NM | 0 | -0.424 | 0.028 | 0.12 | - |
| | | | | | |
| BN_partial_ed_HGQD_308_AFM | 0 | -0.408 | 0 | 0.42 | 0.02 |
| BN_partial_ed_HGQD_308_FM | 2.2 | -0.407 | 0.051 | 0.1 | 0.08 |
| BN_partial_ed_HGQD_308_NM | 0 | -0.407 | 0.114 | 0.05 | - |
| | | | | | |
| BN_partial_ed_HGQD_440_AFM | 0 | -0.399 | 0 | 0.3 | 0.03 |
| BN_partial_ed_HGQD_440_FM | 3.4 | -0.399 | 0.027 | 0.14 | 0.15 |
| BN_partial_ed_HGQD_440_NM | 0 | -0.399 | 0.205 | 0.05 | - |
| | | | | | |
| BN_full_ed_HGQD_176_AFM | 0 | -0.357 | 0.001 | 1.19 | 1.19 |
| BN_full_ed_HGQD_176_FM | 0 | -0.357 | 0.005 | 1.19 | 1.19 |
| BN_full_ed_HGQD_176_NM | 0 | -0.357 | 0 | 1.19 | - |
| | | | | | |
| BN_full_ed_HGQD_308_AFM | 0 | -0.340 | 0.027 | 0.73 | 0.73 |
| BN_full_ed_HGQD_308_FM | 0 | -0.340 | 0 | 0.73 | 0.73 |
| BN_full_ed_HGQD_308_NM | 0 | -0.340 | 0.012 | 0.73 | - |
| | | | | | |
| BN_full_ed_HGQD_440_AFM | 0 | -0.332 | 0.07 | 0.69 | 0.69 |
| BN_full_ed_HGQD_440_FM | 0 | -0.332 | 0.07 | 0.69 | 0.69 |
| BN_full_ed_HGQD_440_NM | 0 | -0.332 | 0 | 0.69 | - |
| | | | | | |
| HBNQD_176_AFM | 0 | -0.460 | 0 | 4.26 | 4.26 |
| HBNQD_176_FM | 0 | -0.460 | 0.002 | 4.26 | 4.26 |
| HBNQD_176_NM | 0 | -0.460 | 0.001 | 4.26 | - |
| | | | | | |
| HBNQD_308_AFM | 0 | -0.446 | 0.005 | 4.14 | 4.14 |
| HBNQD_308_FM | 0 | -0.446 | 0 | 4.14 | 4.14 |
| HBNQD_308_NM | 0 | -0.446 | 0.003 | 4.14 | - |
| | | | | | |
| HBNQD_440_AFM | 0 | -0.439 | 0.003 | 4.05 | 4.05 |
| HBNQD_440_FM | 0 | -0.439 | 0.002 | 4.05 | 4.05 |
| HBNQD_440_NM | 0 | -0.439 | 0 | 4.05 | - |
| | | | | | |
| C_partial_ed_HBNQD_176_AFM | 0 | -0.419 | 0.001 | 0.008 | 0.45 |
| C_partial_ed_HBNQD_176_FM | 1.1 | -0.419 | 0 | 0.64 | 0.60 |
| C_partial_ed_HBNQD_176_NM | 0 | -0.419 | 0.034 | 0.059 | - |
| C_partial_ed_HBNQD_308_AFM | 0 | -0.408 | 0.001 | 0.26 | 0.01 |
| C_partial_ed_HBNQD_308_FM | 0.9 | -0.408 | 0 | 0.12 | 0.14 |
| C_partial_ed_HBNQD_308_NM | 0 | -0.408 | 0.005 | 0.067 | - |
| | | | | | |
| C_partial_ed_HBNQD_440_AFM | 0 | -0.403 | 0.004 | 0.02 | 0.21 |
| C_partial_ed_HBNQD_440_FM | 0.8 | -0.403 | 0.001 | 0.11 | 0.12 |
| C_partial_ed_HBNQD_440_NM | 0 | -0.403 | 0 | 0.072 | - |
| | | | | | |
| C_full_ed_HBNQD_176_AFM | 0 | -0.408 | 0 | 0.80 | 0.80 |
| C_full_ed_HBNQD_176_FM | 0 | -0.408 | 0.001 | 0.80 | 0.80 |
| C_full_ed_HBNQD_176_NM | 0 | -0.408 | 0.002 | 0.80 | - |
| | | | | | |
| C_full_ed_HBNQD_308_AFM | 0 | -0.389 | 0 | 0.48 | 0.48 |
| C_full_ed_HBNQD_308_FM | 0 | -0.389 | 0.006 | 0.48 | 0.48 |
| C_full_ed_HBNQD_308_NM | 0 | -0.389 | 0.016 | 0.48 | - |
| | | | | | |
| C_full_ed_HBNQD_440_AFM | 0 | -0.379 | 0.002 | 0.25 | 0.25 |
| C_full_ed_HBNQD_440_NM | 0 | -0.379 | 0 | 0.25 | - |

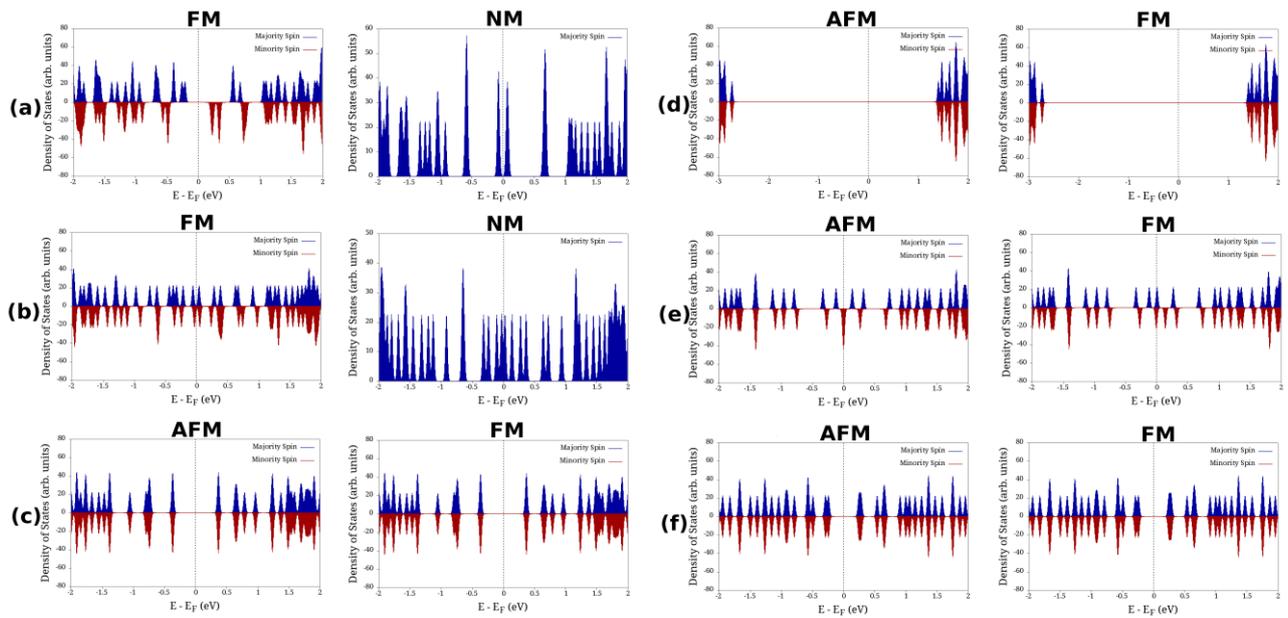

Figure S2: DOS plots of all the other spin configurations of (a) HGQD, (b) BN-partial-ed-HGQD, (c) BN-full-ed-HGQD, (d) HBNQD, (e) C-partial-ed-HBNQD and (f) C-full-ed-HBNQD of size (4.39, 4.16)